\documentclass[aps,prl,reprint,notitlepage,showpacs,nofootinbib,superscriptaddress,twocolumn%,raggedbottom
,nofloats,preprintnumbers]{revtex4-1}
\usepackage{lineno}
\usepackage{graphicx}
\usepackage[utf8]{inputenc} 
\usepackage{amsmath}
\usepackage{amssymb}
\usepackage{ulem}
\usepackage{float}
\usepackage{comment}
\usepackage{slashed}
\usepackage{subfigure}

\usepackage[T1]{fontenc} % if needed
\usepackage{parskip}
\usepackage{bm}         % for bold math characters in a math environment
\usepackage{array}
\usepackage{multirow}   % analogue to \multicolumn
\usepackage{booktabs}   % for fancy looking tables

\usepackage{textcomp}
\usepackage{amssymb}
\usepackage{amsmath}
\usepackage{soul}

\usepackage{enumerate}
\usepackage{caption}
\usepackage{subcaption}
\usepackage{ragged2e}
 \DeclareCaptionJustification{justified}{\justifying}
\usepackage{amsmath}
\usepackage{amssymb}
\usepackage{indentfirst}
\usepackage[usenames,dvipsnames]{xcolor}
\usepackage[colorlinks=true,citecolor=darkred,urlcolor=darkred, pdfborder={0 0 0}]{hyperref}
\definecolor{darkred}{rgb}{0.6,0,0}

\definecolor{linkcolor}{rgb}{0,0,0.5}
\newcommand {\ignore}[1]{}

\definecolor{bostonuniversityred}{rgb}{0.8, 0.0, 0.0}

\def\gsim{\raise0.3ex\hbox{$\;>$\kern-0.75em\raise-1.1ex\hbox{$\sim\;$}}}
\def\lsim{\raise0.3ex\hbox{$\;<$\kern-0.75em\raise-1.1ex\hbox{$\sim\;$}}}

\usepackage{soul}
\definecolor{mightnightblue}{RGB}{25,25,112}

\definecolor{brown}{rgb}{0.59, 0.29, 0.0}

\newcommand{\eps}{\varepsilon}

\def\21{$\mathrm{SU(2)_L \otimes U(1)_Y}$}

\newcommand{\AddrUNAM}{ {\it Instituto de F\'{\i}sica, Universidad Nacional Aut\'onoma de M\'exico, A.P. 20-364, Ciudad de M\'exico 01000, M\'exico.}}

\newcommand{\AddrLJF}{ {\it Tecnol\'ogico Nacional de M\'exico/ITS de Jerez, C.P. 99863, Zacatecas, M\'exico.}}

\begin{document}

\title{\boldmath\color{BrickRed} Exploring the Standard Model and Beyond from the Evidence of CE$\nu$NS with Reactor Antineutrinos in CONUS+}
\author{M.~Alpízar-Venegas}\email{marioalpizarv@estudiantes.fisica.unam.mx }
\affiliation{\AddrUNAM}
%\author{L. M. Garcia de la Vega}\email{leonm@ciencias.unam.mx}\affiliation{\AddrUNAM}

\author{L. J. Flores}\email{ljflores@jerez.tecnm.mx }
\affiliation{\AddrLJF}

\author{Eduardo Peinado}\email{epeinado@fisica.unam.mx}
\affiliation{\AddrUNAM}

\author{E.~V\'azquez-J\'auregui}\email{ericvj@fisica.unam.mx }
\affiliation{\AddrUNAM}
%\affiliation{Department of Physics, Laurentian University, Sudbury, P3E 2C6, Canada}

\vspace{0.7cm}

\begin{abstract}
\vspace{0.3cm}

The observation of the Coherent Elastic Neutrino-Nucleus Scattering (CE$\nu$NS) process using reactor antineutrinos offers a unique opportunity to probe the Standard Model and explore Beyond the Standard Model scenarios. This study reports on the latest results from the CONUS+ experiment conducted at the Leibstadt nuclear power plant in Switzerland. The CONUS collaboration reports $395 \pm 106$ events detected from reactor antineutrinos with an exposure of 327 kg$\cdot$days, utilizing high-purity germanium detectors operated at sub-keV thresholds. A $\chi^2$-based statistical analysis was performed on these results, incorporating systematic uncertainties. This analysis was used to extract the weak mixing angle, establish a limit on the neutrino magnetic moment, and impose constraints on neutrino non-standard interactions using reactor antineutrinos. The results confirm the potential of CE$\nu$NS experiments in the study of fundamental neutrino properties and probing new physics.

\end{abstract}

\maketitle

\subsection{Introduction}

\noindent

 Another piece of evidence for the observation of the Coherent Elastic neutrino-Nucleus Scattering (CE$\nu$NS) process for reactor antineutrinos~\cite{Ackermann:2025obx} is an opportunity to explore the Standard Model (SM) and several Beyond the Standard Model (BSM) scenarios. Measurements have been reported using semiconductor detectors made of high-purity germanium crystals operated at sub-keV energy thresholds. The detector was placed near a boiling water reactor in the Leibstadt nuclear power plant (KKL), Switzerland. 

The first observation of the CE$\nu$NS process was performed in a pion-at-rest decay source in 2017 by the COHERENT collaboration using a CsI[Na] crystal \cite{COHERENT:2017ipa}, followed by a measurement using a LAr detector \cite{COHERENT:2020iec} in 2020 at the Spallation Neutron Source SNS at Oak Ridge National Laboratory. The COHERENT collaboration also reported the first detection on germanium \cite{COHERENT:2024axu}, employing high-purity germanium spectrometers operated at an energy threshold of 1.5 keV.

Measuring CE$\nu$NS with reactor antineutrinos offers some advantages compared to the pion-at-rest decay source. The energy of neutrinos from a nuclear reactor is below 10 MeV, which means that the form factor is nearly one at these energies. In addition, reactor neutrinos consist of purely electron antineutrinos, offering the possibility to explore antineutrino disappearance in search of sterile neutrinos \cite{Alfonso-Pita:2022eli}. 

In addition, this process was recently observed from solar neutrinos interacting with dedicated experiments to search for dark matter such as PandaX-4T \cite{PandaX:2024muv} and XENONnT \cite{XENON:2024ijk}. Both experiments measured the solar ${^8}$B neutrino flux by detecting neutrinos through coherent scattering with xenon nuclei.

All these measurements open a window of opportunity to continue probing the Standard Model, measuring several properties and parameters, such as the weak mixing angle or the neutrino magnetic moment~\cite{Majumdar:2022nby, Maity:2024aji, Demirci:2024vzk, DeRomeri:2024hvc}. Furthermore, it can also test new physics, such as the existence of non-standard interactions (NSI) of the neutrinos~\cite{AristizabalSierra:2022axl, AristizabalSierra:2024nwf,Li:2024iij, DeRomeri:2024iaw,Blanco-Mas:2024ale}. This work explores different physics scenarios from the latest results of the CONUS+ experiment on the recent observation of reactor antineutrinos through coherent scattering. 
 
\subsection{Experiment and Analysis Description}

CONUS+~\cite{CONUS:2024lnu} is an experiment conducted at the Leibstadt nuclear power plant in Switzerland. The reactor has a thermal power of 3.6 GW. The experimental setup is located at 20.7 m from the core and consists of four high-purity germanium (HPGe) detectors with a total fiducial mass of 3.73 $\pm$ 0.02 kg. This experiment is the successor of the CONUS~\cite{CONUSCollaboration:2024oks} experiment that operated at the Brokdorf Nuclear Power Plant in Germany.

The results presented in~\cite{Ackermann:2025obx} were obtained by using three of the four detectors, each with different operating energy thresholds of 160, 170, and 180 eV, respectively. The data set employed in the analysis has an exposure of 327 kg per day, acquired from November 2023 to July 2024. Cosmogenic and radiogenic backgrounds were measured with dedicated ancillary systems, evaluating contributions of neutrons and gammas induced by muons~\cite{CONUS:2024vyx}. These measurements were important due to the shallow overburden where the detector is located ($7.4$ meters equivalent water). The background measured was subtracted from the signal observed.

The study reported in this manuscript assumes the Huber-Muller spectrum normalized to the neutrino flux reported in~\cite{Ackermann:2025obx}. This flux is $1.5\times10^{13}$anti$\nu/(cm^2s)$. The response of the detectors to nuclear recoils was described using the individual trigger efficiency for each detector. These functions were determined at the experimental site by the CONUS collaboration for the four detectors~\cite{CONUS:2024lnu}. Moreover, the Lindhard theory was applied with a parameter $k=0.162 \pm 0.004$ (error includes both statistical and systematic uncertainties)~\cite{Bonhomme:2022lcz} to obtain the ionization energy. The data reported from CONUS+, as well as previous CONUS result~\cite{CONUSCollaboration:2024oks} and~\cite{Bonhomme:2022lcz},
do not indicate deviations from this semi-empirical model. The full exposure was calculated using the individual energy thresholds (160, 170, and 180 eV), livetimes (110, 119, and 117 days), and masses (0.94, 0.94, and 0.95 kg) for each detector. 
This analysis resulted in $330$ events predicted by the Standard Model.

The Standard Model cross-section for CE$\nu$NS  is 
\begin{equation}
\frac{d\sigma}{dT} = \frac{G_F^2}{2\pi}M_N Q_w^2 \left(2 - \frac{M_N T}{E_\nu^2}\right).
\label{eq:crossSec}
\end{equation}
Here, $m_N$, $Z$, and $N$ are the nuclear mass, proton, and neutron number of the detector material respectively. In Eq. (\ref{eq:crossSec}), the weak charge $Q_w$ is given by $Q_w = Z g_p^V F_Z(q^2) + N g_n^V F_N(q^2)$, where $F_Z$ and $F_N$ are the proton and neutron form factors. A fit with the following $\chi^2$ function was performed

\begin{equation}
    \chi^2 =\underset{\alpha}{\min}\left[\left(\frac{N_\mathrm{meas} - (1+\alpha)N_\mathrm{th}(X)}{\sigma_\mathrm{stat}} \right)^2 + \left(\frac{\alpha}{\sigma_\alpha}\right)^2 \right]
    \label{eq:chisq}
\end{equation}
where $N_\mathrm{meas}$ is the number of events measured, $N_\mathrm{th}(X)$ is the theoretical prediction for the number of events, $\sigma_\mathrm{stat} = \sqrt{N_\mathrm{meas}}$ is the statistical uncertainty, and $\sigma_{\alpha}$ includes the systematic uncertainties from the energy threshold (14.1\%), quenching parameter (7.3\%), reactor neutrino flux (4.6\%), cross-section (3.2\%), active germanium mass (1.1\%), trigger efficiency (0.7\%), and a systematic uncertainty of 21.8\% accounting for the likelihood fit, fit method, background model, and non-linearity implementation obtained from the signal reported by the CONUS collaboration. The variable $X$ refers to the parameter to be fitted. The $\chi^2$ function is minimized over the nuisance parameter $\alpha$. 

The following sections describe the extraction of the weak mixing angle, the calculation of a limit to the neutrino magnetic moment, and the constraints on NSI parameters.

\subsection{The Weak Mixing Angle}

\begin{figure}[t]
    \centering
    \includegraphics[width=\linewidth]{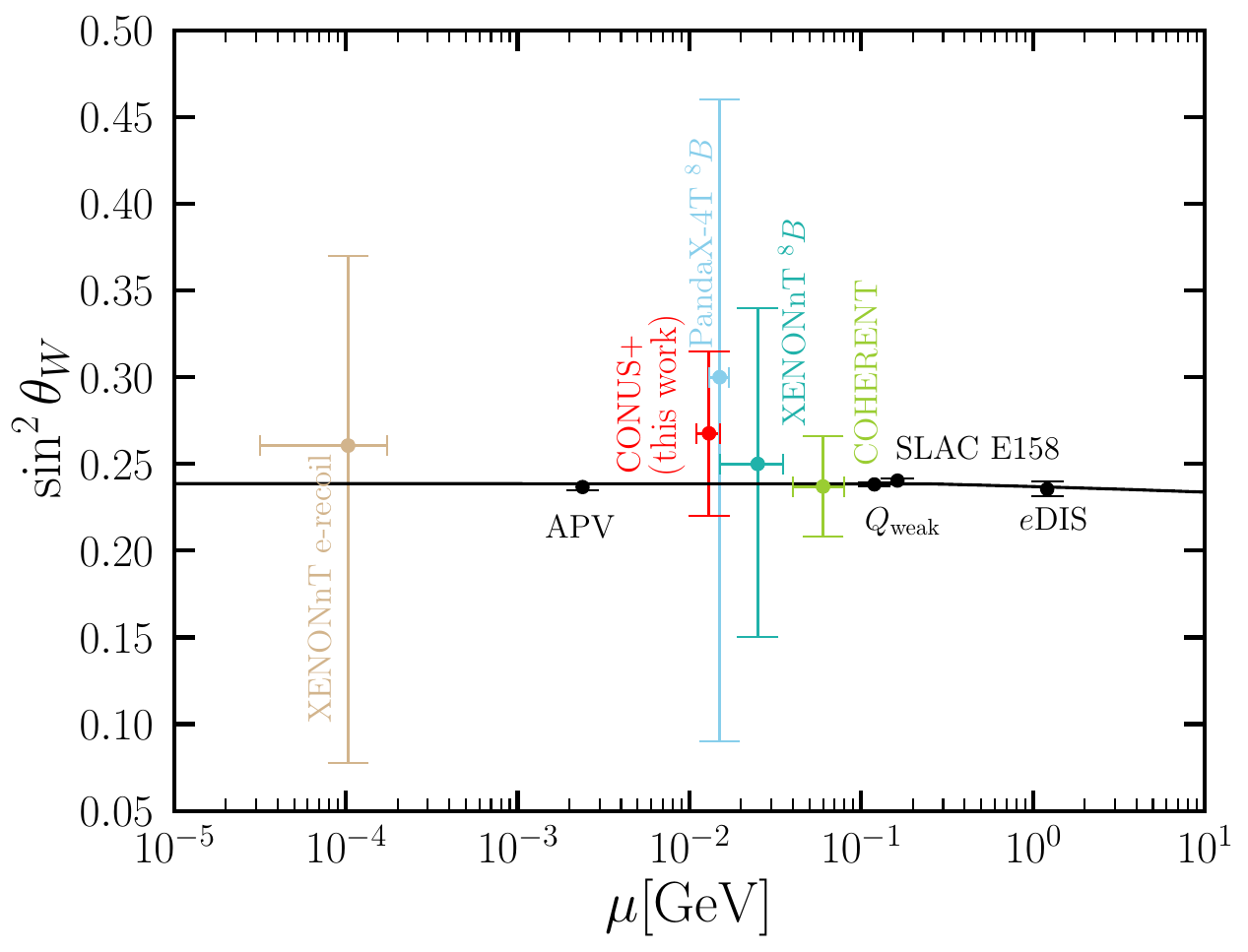}
    \caption{RGE running of the weak mixing angle in the $\overline{\mathrm{MS}}$ renormalization scheme as a function of the energy scale $\mu$~\cite{Erler:2017knj,ParticleDataGroup:2024cfk}. The $1\sigma$ determination from the CONUS+ measurement is shown in red. Measurements from different experiments are also presented~\cite{Antypas_2018, Prescott:1979dh, SLACE158:2005uay, Qweak:2018tjf, ParticleDataGroup:2024cfk, DeRomeri:2022twg, Maity:2024aji, DeRomeri:2024iaw}. 
      }
    \label{fig:weakMixingAngle}
\end{figure}

The extraction of the weak mixing angle is one of the most important electroweak precision tests of the Standard Model \cite{Erler:2017knj,ParticleDataGroup:2024cfk}. Several experiments have determined it at different energy scales, and CE$\nu$NS gives an opportunity to determine this angle at low momentum transfer as a complement to the one provided by atomic parity violation (APV)~\cite{Antypas_2018} and Moller scattering $Q_w$~\cite{SLACE158:2005uay}. A fit can be performed to $\sin^2\theta_W$ from Eq.~\eqref{eq:chisq}. Since the momentum transfer is very low in reactor neutrinos, the form factors of neutron and proton are very close to unit. If no new physics contribution exists to the CE$\nu$NS differential cross-section, the weak mixing angle can be extracted through the SM weak coupling $g_p^V = 1/2 - 2 \sin^2\theta_W$. 
Eq.~\eqref{eq:chisq} is used to fit the weak mixing angle by computing the theoretical number of events as a function of $\sin^2\theta_W$. In Fig.~\ref{fig:weakMixingAngle}, the $1\sigma$ determination from the CONUS+ measurement is shown, along the RGE running of the weak mixing angle as a function of the energy scale in the $\overline{\mathrm{MS}}$ renormalization scheme~\cite{Erler:2017knj,ParticleDataGroup:2024cfk}.  The result obtained for the weak mixing angle is sin$^{2}\theta_{W}$ = 0.268 $\pm$ 0.047, which is consistent with the findings of the other CE$\nu$NS experiments presented. Other measurements are presented, from APV~\cite{ParticleDataGroup:2024cfk, Antypas_2018},  elastic electron-proton scattering (Q$_\mathrm{weak}$)\cite{Qweak:2018tjf}, Moller scattering (SLAC E158)\cite{SLACE158:2005uay}, electron-deuteron deep inelastic scattering (eDIS)\cite{Prescott:1979dh}, neutrino-electron scattering (XENONnT)\cite{Maity:2024aji}, CE$\nu$NS from spallation neutron source (COHERENT CsI \& LAr)\cite{DeRomeri:2022twg} and from solar neutrinos (PANDAX-4T \& XENONnT)\cite{DeRomeri:2024iaw}. Results from experiments using neutrinos still have considerably higher uncertainties when compared to measurements from experiments using electrons, as seen in Fig.~\ref{fig:weakMixingAngle}.

\subsection{The Neutrino Magnetic Moment}

The neutrino magnetic moment has been widely studied for massive Majorana and Dirac neutrinos~\cite{Kayser:1981nw,Vogel:1989iv,Giunti:2014ixa,Fujikawa:1980yx,Pal:1981rm,Shrock:1982sc,Dvornikov:2003js,Dvornikov:2004sj, Barr:1990um, Babu:1990wv,Pal:1991qr,Boyarkin:2014oza,Giunti:2024gec} and has been constrained by different types of experiments~\cite{Vidyakin:1992nf,Derbin:1993wy,MUNU:2005xnz,TEXONO:2006xds,Beda:2012zz,CONUS:2022qbb,Allen:1992qe,Ahrens:1990fp,LSND:2001akn,Cooper-Sarkar:1991vsl,DONUT:2001zvi,Coloma:2022avw,AtzoriCorona:2022qrf,Grotch:1988ac,Tanimoto:2000am}. 
The non-zero neutrino magnetic moment induces a new interaction that contributes to the CE$\nu$NS cross-section without interference as 
\begin{equation}
    \frac{d\sigma}{dT} = \pi\frac{\alpha_{\mathrm{EM}}^2Z^2 \mu_\nu^2}{m_e^2} \left(\frac{1}{T} - \frac{1}{E_\nu} + \frac{T}{4E_\nu^2}  \right) F^2(q^2).
    \label{eq:mu_nu}
\end{equation}
Here, the neutrino magnetic moment $\mu_\nu$ is normalized by the Bohr magneton $\mu_B$, $m_e$ is the electron mass, and $\alpha_{\mathrm{EM}}$ is the electromagnetic fine structure constant.
\begin{figure}[t]
    \centering
    \includegraphics[width=\linewidth]{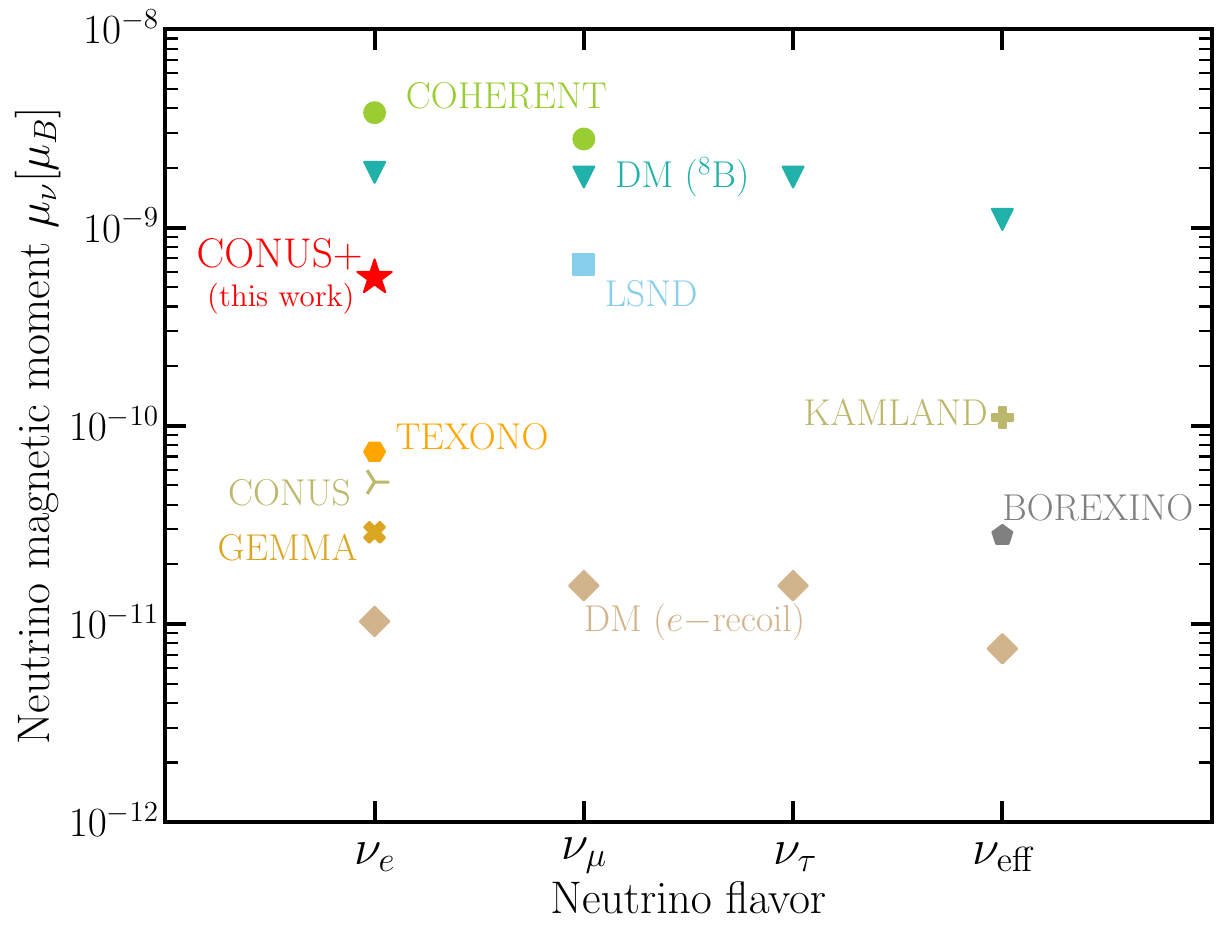}
    \caption{90\% C.L. limits for the neutrino magnetic moment. The results from this analysis of the CONUS+ experiment are competitive with existing limits derived from nuclei scattering measurements~\cite{DeRomeri:2022twg}, including dark matter direct detection experiments~\cite{DeRomeri:2024hvc}. The most stringent limits come from dark matter direct detection experiments extracted from electron scattering~\cite{Giunti:2023yha}. Other limits derived from different experiments are also presented for comparison~\cite{LSND:2001akn, Super-Kamiokande:2004wqk,TEXONO:2006xds, Beda:2012zz, Borexino:2017fbd,Buck:2024bkh}
    }
    \label{fig:magneticMoment}
\end{figure}

The limits from the $\chi^2$ analysis to the CONUS+ results are presented in Fig.~\ref{fig:magneticMoment}. Results from direct dark matter
detection experiments through elastic neutrino-electron
scattering~\cite{Giunti:2023yha} and coherent elastic neutrino-nucleus scattering~\cite{DeRomeri:2024hvc} are also shown. COHERENT has also placed limits using CsI and LAr~\cite{DeRomeri:2022twg}. The bound in the neutrino magnetic moment obtained from the CONUS+ results is $\mu_\nu$=$5.6\times10^{-10} \mu_B$ at 90\% C.L. This limit is less stringent than those obtained from electron scattering in dark matter direct detection experiments and approximately an order of magnitude higher than the constraints placed by other reactor experiments, such as TEXONO~\cite{TEXONO:2006xds} and GEMMA~\cite{Beda:2012zz}. Previous results from CONUS set a more stringent constraint, $\mu_\nu =5.2\times10^{-11} \mu_B$ at 90\% C.L.~\cite{Buck:2024bkh}, due to a significantly larger energy window and exposure compared to the CONUS+ results analyzed in this work.

\subsection{Non-standard Interactions}

\begin{figure}[H]
    \centering
    \includegraphics[width=\linewidth]{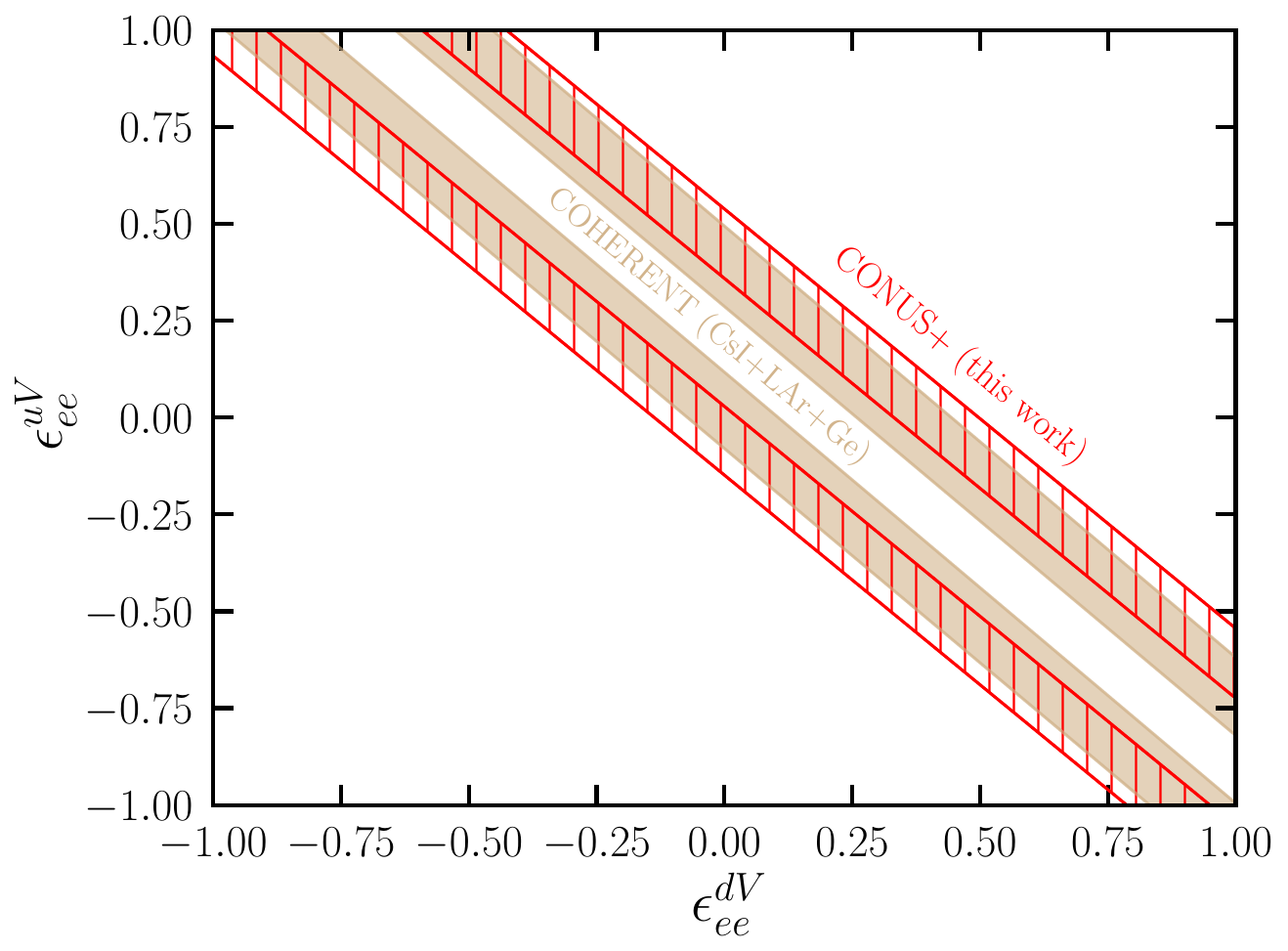}
    \caption{Predicted 95\% C.L. sensitivity obtained from the analysis of the CONUS+ results. Allowed parameters established by the COHERENT experiment using CsI, LAr, and Ge results~\cite{Liao:2024qoe} are also shown.}
    \label{fig:NSI}
\end{figure}

Any deviation from the SM CE$\nu$NS cross-section would modify the expected number of events, hinting at physics beyond the Standard Model. The non-standard interactions (NSI) formalism is a way to parametrize the possible new physics that consists of modifying the neutral current component with the extra contribution~\cite{Miranda:2015dra}:
\begin{equation}
    \mathcal{L}_\mathrm{NC}^\mathrm{NSI} = -2\sqrt{2}G_F \sum_{f,P,\alpha,\beta}  \eps_{\alpha\beta}^{fP}( \bar{\nu}_\alpha \gamma^\mu P_L \nu_\beta)   (  \bar{f}\gamma_\mu P_X f).\label{NSIint}
\end{equation}
In Eq. (\ref{NSIint}), $f$ represents $u$ and $d$ quarks, $\alpha$ and $\beta$ correspond to the neutrino flavors $(e, \mu, \tau)$, $P_X$ is the chirality projectors $L$ and $R$, and $\eps_{\alpha\beta}^{fP}$ are the couplings  characterizing the strength of the NSI. Fig. \ref{fig:NSI} presents the limits set by this analysis to the CONUS+ results. The results, reported at 95\% C.L., indicate similar sensitivity to the combination of the results from the COHERENT collaboration using caesium iodide (CsI), liquid argon (LAr), and germanium (Ge)~\cite{Liao:2024qoe}.

\subsection{Conclusions}

The CE$\nu$NS process is a powerful tool for placing constraints on both the Standard Model and potential BSM physics. An analysis of the CONUS+ experiment results, recently reported in~\cite{Ackermann:2025obx}, has provided a limit on the weak mixing angle, marking an important contribution to the measurement of electroweak parameters at low energies. In particular, this limit represents the first constraint on the weak mixing angle for electron antineutrinos interacting via the CE$\nu$NS process, in a reactor experiment reporting a measurement with a significance greater than 3$\sigma$, providing convincing evidence for this phenomenon. The bound on the neutrino magnetic moment is also competitive, especially when compared to direct dark matter detection experiments scattering nuclei and accelerator CE$\nu$NS experiments, such as COHERENT. The constraints imposed by TEXONO and GEMMA are still leading the results for experiments measuring reactor neutrinos. Lastly, the constraints on the NSI parameters $\epsilon_{ee}^{dV}-\epsilon_{ee}^{uV}$ from CONUS+ are similar to those coming from the COHERENT collaboration data.\\
These results also demonstrate the sensitivity of CE$\nu$NS to new physics, offering a strong framework for exploring extensions of the Standard Model and further improving our understanding of neutrino interactions and their role in fundamental physics. Reducing systematic uncertainties combined with larger datasets will provide stronger constraints on many electroweak parameters and explore BSM with higher precision.

\section{Acknowledgements}
\begin{acknowledgments}
\indent This work is supported by DGAPA UNAM grants PAPIIT-IN105923, PAPIIT-IN111625, and PAPIIT-IA104223. EP and EVJ are grateful for the support of PASPA-DGAPA, UNAM for a sabbatical leave, and Fundación Marcos Moshinsky.

\end{acknowledgments}
\bibliographystyle{apsrev4-1}
\bibliography{bibliography}
\end{document}